\documentclass{article}

\usepackage[preprint]{neurips_2026}

\usepackage[utf8]{inputenc} 
\usepackage[T1]{fontenc}    
\usepackage{hyperref}       
\usepackage{url}            
\usepackage{booktabs}       
\usepackage{amsfonts}       
\usepackage{nicefrac}       
\usepackage{microtype}      
\usepackage{xcolor}         
\usepackage{graphicx} 

\usepackage{float}        

\usepackage{algorithm}
\usepackage{algpseudocode}

\usepackage{amsmath}
\usepackage{enumitem}
\usepackage{multirow}

\title{MemFlow: Intent-Driven Memory Orchestration for Small Language Model Agents}

\author{%
  Jiayi Chen \quad Yingcong Li \quad Guiling Wang \\
  New Jersey Institute of Technology, Newark, USA \\
  \texttt{\{jc2693, yingcong.li, guiling.wang\}@njit.edu}
}

\begin{document}

\maketitle

\begin{abstract}
Modern language agents must operate over long-horizon, multi-turn histories, yet deploying such agents with Small Language Models (SLMs) remains fundamentally difficult. Full-context prompting causes context overflow, flat retrieval exposes the model to noisy evidence, and open-ended agentic loops are unreliable under limited reasoning capacity. We argue that a substantial portion of SLM memory failure arises from mismatched memory operations: different query types demand categorically different retrieval strategies, evidence transformations, and context budgets that SLMs cannot reliably self-orchestrate through open-ended reasoning. We introduce \textit{MemFlow}, a training-free memory orchestration framework that externalizes memory planning from the SLM. A Router Agent classifies each query by intent and dispatches it to the Memory Agent, which executes one of three specialized tiers (Profile Lookup, Targeted Retrieval, or Deep Reasoning) and assembles the resulting evidence under a dynamic, tier-aware token budget. An Answer Agent then generates a response from this compact context, and a Validator Agent optionally retries with a heavier memory tier when the response is not supported by the provided evidence. This route-then-compile design avoids tool-selection hallucination and reasoning loops while keeping the answer context compact.
Evaluated on a frozen Qwen3-1.7B backbone across long-horizon memory benchmarks---LongMemEval, LoCoMo, and LongBench---\textit{MemFlow} improves accuracy by nearly $2\times$ over full-context SLM baselines. These results suggest that structured intent routing and deterministic evidence preparation can make limited-capacity models substantially more effective in resource-constrained long-horizon agents.
\end{abstract}


\section{Introduction}

Language agents operating over long-horizon interactions must answer queries
whose evidence spans many turns, sessions, and topic shifts~\cite{wu2024longmemeval,maharana2024locomo,packer2023memgpt,park2023generative}. The common strategy of appending the full history causes unbounded memory growth, high cost, and degraded reasoning once inputs exceed the model's effective attention span~\cite{liu2024lost}. Frontier LLMs partly absorb this cost with scale and extended context windows~\cite{openai2024gpt4o,anthropic2024claude,google2024gemini}, but Small Language Models (SLMs) face hard context limits in resource-constrained deployments~\cite{abdin2024phi3,liu2024mobilellm,qwen2025qwen3,allal2025smollm2,zhou2025pruningslm}.

RAG~\cite{lewis2020rag,karpukhin2020dpr} mitigates overflow by selecting evidence, but uniform retrieval cannot serve the structural diversity of long-horizon queries~\cite{wu2024longmemeval,bai2024longbench}: preferences, timelines, knowledge updates, and multi-session synthesis require different retrieval strategies, transformations, and budgets. ReAct-style reasoning~\cite{yao2023react} and learned tool use~\cite{schick2023toolformer} provide flexibility, but sub-3B agents frequently produce hallucinated calls or broken reasoning traces~\cite{faghih2025tool}. We therefore frame a substantial portion of SLM memory failure as an \emph{intent-routing mismatch}: for SLM agents, the central question is not only what to retrieve, but which memory operation the query requires.

\begin{figure}[t]
    \centering
    \includegraphics[width=\textwidth]{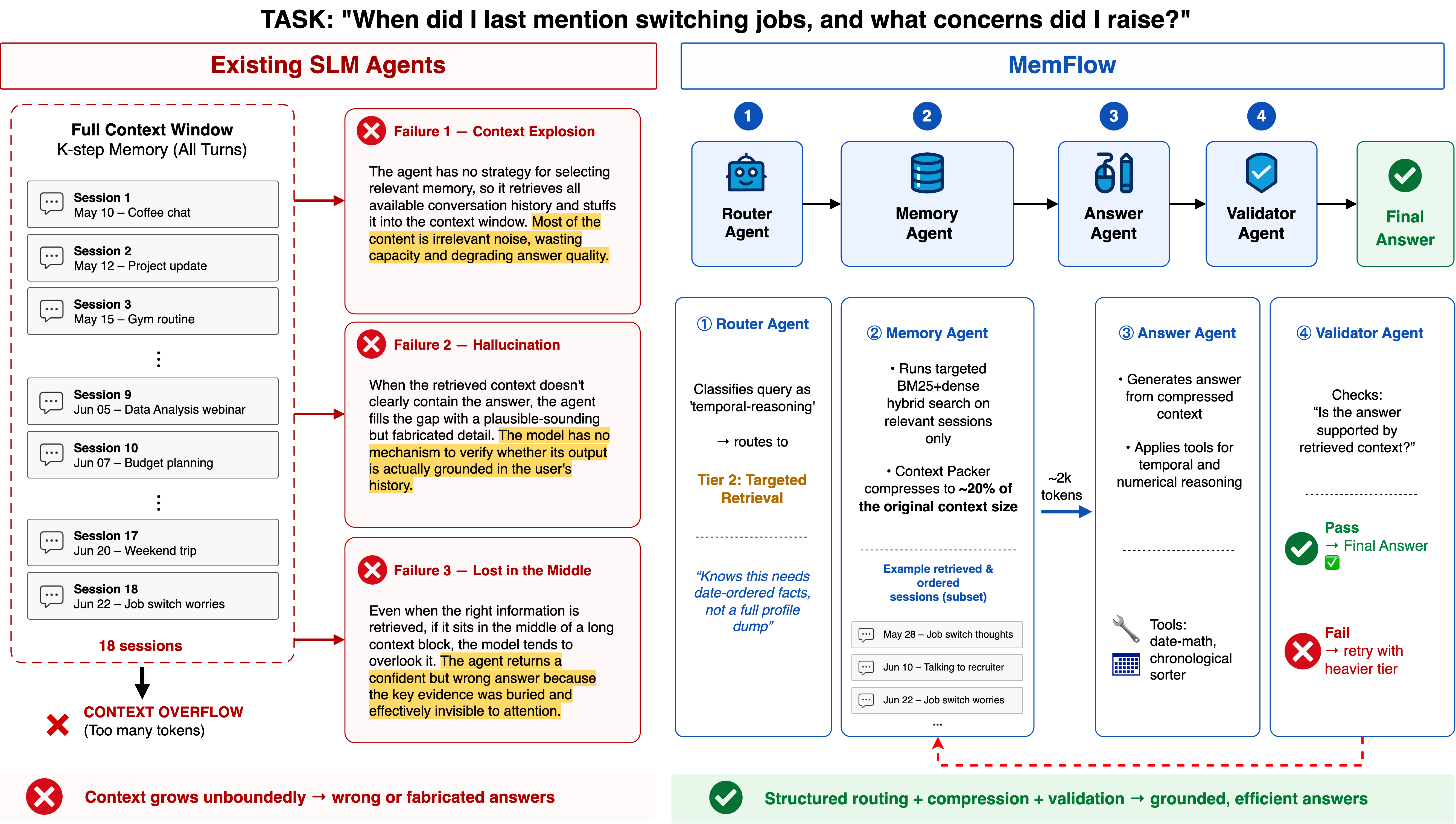}
    \caption{Comparison of existing SLM memory approaches and \textit{MemFlow}.
    \textbf{Left:} existing agents suffer from context explosion,
    hallucination, and lost-in-the-middle failures when full histories or
    uniform retrieval are used with open-ended reasoning loops.
    \textbf{Right:} \textit{MemFlow} addresses these failure modes through
    intent-driven routing to specialized memory tiers, priority-aware context
    compilation under a dynamic tier-aware token budget, and grounding-validated escalation.}
    \label{fig:memflow_comparison}
\end{figure}

Prior work addresses pieces of this problem. Prompt compressors such as LongLLMLingua~\cite{jiang2023longllmlingua} and LLMLingua-2~\cite{pan2024llmlingua2} reduce context but are task-agnostic; Self-RAG~\cite{asai2023selfrag} and CRAG~\cite{yan2024crag} add critique but still delegate orchestration to the model. Memory systems such as MemGPT~\cite{packer2023memgpt}, Mem0~\cite{chhikara2025mem0}, and Zep~\cite{rasmussen2025zep} provide external stores, while routing methods~\cite{ong2024routellm,saha2024system} allocate compute across models or subgoals. MEM1~\cite{zhou2025mem1} learns compact memory end-to-end, but requires task-specific reinforcement learning. A training-free framework that jointly handles intent classification, retrieval specialization, context budgeting, and grounding validation for frozen SLMs remains missing.

We introduce \textit{MemFlow}, a training-free, intent-driven memory orchestration framework for frozen SLM agents (Figure~\ref{fig:memflow_comparison}). A \textbf{Router Agent} classifies each query and dispatches it to a \textbf{Memory Agent} tier: direct profile lookup, targeted retrieval, or deep reasoning with deterministic preprocessing for temporal sorting, conflict resolution, and synthesis. The Memory Agent packs evidence under a dynamic, priority-aware token budget; an \textbf{Answer Agent} generates from this compact context; and a \textbf{Validator Agent} triggers heavier-tier recovery on grounding failure. This \emph{route-then-compile} design reduces open-ended orchestration: after routing, execution follows a fixed memory path and the answer model sees intent-matched evidence rather than raw history.
Across LongMemEval~\cite{wu2024longmemeval}, LoCoMo~\cite{maharana2024locomo}, and LongBench~\cite{bai2024longbench}, MemFlow nearly doubles full-context accuracy with a frozen Qwen3-1.7B~\cite{qwen2025qwen3} and improves over RAG/ReAct by $+6.2$\,pp while keeping the final answer context compact. Because the full orchestration pipeline is costlier than the final context alone, we report both full-pipeline cost and answer-context size. MemFlow also improves additional sub-3B backbones---Qwen3-0.6B~\cite{qwen2025qwen3}, SmolLM2-1.7B~\cite{allal2025smollm2}, LLaMA-3.2-1B~\cite{meta2024llama32}, and Gemma-3-1B~\cite{gemma2025gemma3}---suggesting that structured memory routing can substitute for some missing model capacity without training.

\section{Related Work}

\textbf{Retrieval, compression, and evidence preparation.}
RAG~\cite{lewis2020rag,karpukhin2020dpr,guu2020realm} established the retrieved-memory paradigm; FiD~\cite{izacard2020fid}, Contriever~\cite{izacard2021contriever}, and ColBERTv2~\cite{santhanam2022colbertv2} improved passage fusion and retrieval precision. Later systems refine retrieval decisions through reasoning state or critique, including IRCoT~\cite{trivedi2023ircot}, Active RAG~\cite{jiang2023active}, Self-RAG~\cite{asai2023selfrag}, CRAG~\cite{yan2024crag}, and RAFT~\cite{zhang2024raft}; hierarchical and ranking methods such as RAPTOR~\cite{sarthi2024raptor}, GraphRAG~\cite{edge2024graphrag}, and RankRAG~\cite{yu2024rankrag} improve synthesis under limited budgets. Compression and long-context studies show why selective evidence matters: attention underuses middle-positioned evidence~\cite{liu2024lost}; window-extension methods reduce attention cost~\cite{beltagy2020longformer,zaheer2020bigbird,dao2022flashattention,zhang2023h2o}; and LongLLMLingua~\cite{jiang2023longllmlingua} and LLMLingua-2~\cite{pan2024llmlingua2} compress prompts. These systems improve retrieval or compression, but generally apply policies independent of the memory operation a query structurally requires.

\textbf{Memory-augmented language agents.}
Generative Agents~\cite{park2023generative} established persistent memory and reflection for long-horizon agents. MemGPT~\cite{packer2023memgpt} uses OS-style virtual memory paging, Mem0~\cite{chhikara2025mem0} uses a dual vector/graph store, and Zep~\cite{rasmussen2025zep} adds temporal graph structure, but these systems still rely on model-led memory access or similarity-driven retrieval. Reflexion~\cite{shinn2023reflexion} stores episodic failure traces for self-improvement, while MEM1~\cite{zhou2025mem1} learns a compact memory state through reinforcement learning. Recent memory-selection systems also use typed or intent-aware memory, including ENGRAM's lightweight typed stores~\cite{patel2025engram} and MemGuide's intent-aligned multi-session dialogue retrieval~\cite{du2025memguide}. MemFlow instead targets frozen SLMs with bounded route-then-compile execution that couples intent routing, deterministic evidence preparation, and grounding validation.

\textbf{Agentic orchestration, tool reliability, and routing.}
Open-ended orchestration remains brittle: AutoGen~\cite{wu2024autogen} scales multi-agent reasoning, WebArena~\cite{zhou2024webarena} exposes stability failures, and tool-use studies report hallucinated calls and mismatched arguments~\cite{patil2023gorilla,faghih2025tool}. Routing systems such as FrugalGPT~\cite{chen2023frugalgpt}, RouteLLM~\cite{ong2024routellm}, System-1.x~\cite{saha2024system}, and homogeneous-tool query routing for RAG~\cite{mu2024queryrouting} separate dispatch from solving, but mainly route across models, tools, or planning modes. MemFlow applies routing to the memory operations themselves, coupling intent classification with query-type-aware retrieval, context compilation, and grounding validation without additional model training.

\section{MemFlow}
\label{sec:memflow}

\begin{figure}[t]
    \centering
    \includegraphics[width=\textwidth,trim=0 0 0 0, clip]{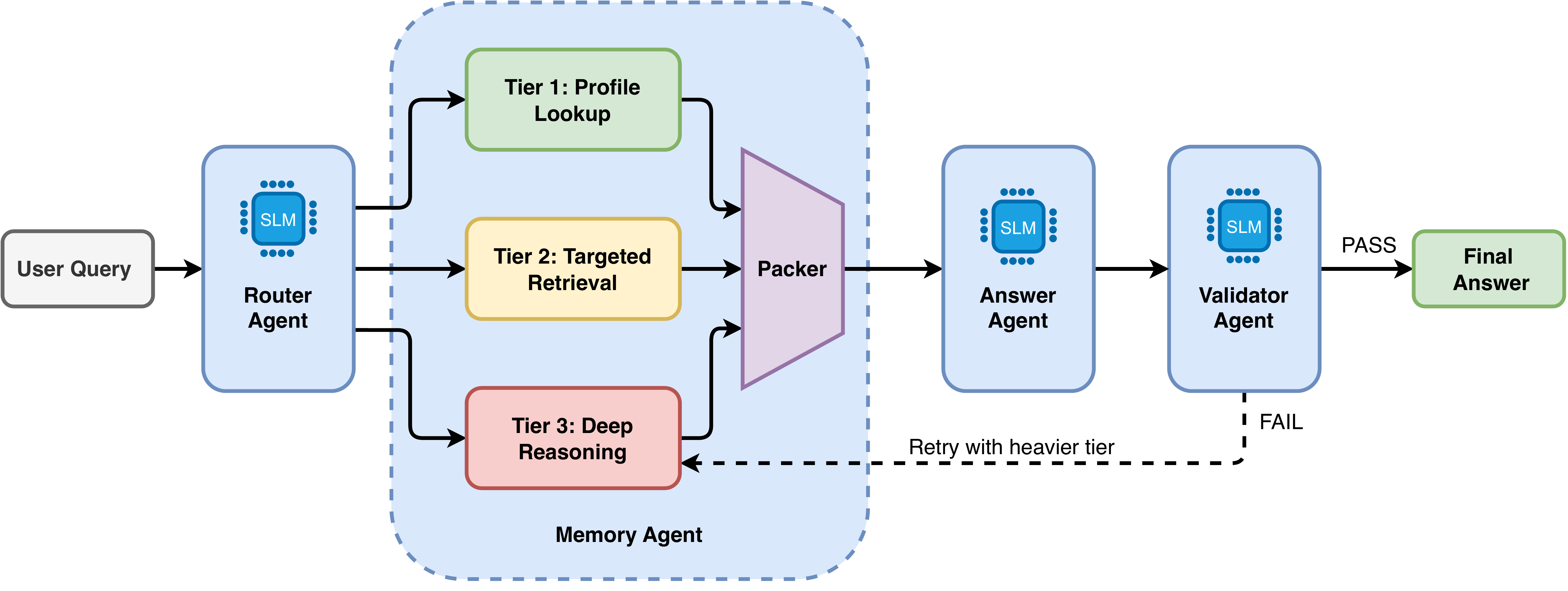}
    \caption{Overview of the \textit{MemFlow} pipeline. 
    SLM chip icons denote SLM inference points.}
    \label{fig:architecture}
\end{figure}

The central premise of MemFlow is that long-horizon memory is not one
problem but a family of structurally distinct problems. A query asking for
a user's dietary preference needs no retrieval at all; a query about the
chronological order of two events requires date arithmetic over retrieved
evidence; a query about the most recent rule governing a constraint needs
post-filtering for policy language. Applying a single retrieval strategy
uniformly across these cases systematically fails on the cases it was not
designed for. MemFlow addresses this by treating memory as an
\emph{intent-conditioned orchestration} problem and externalizing all
memory decisions from the SLM into a structured multi-agent pipeline composed of four components: a \textbf{Router Agent}
that classifies each query by intent, a \textbf{Memory Agent} that
executes the corresponding retrieval and evidence compilation, an
\textbf{Answer Agent} that generates a grounded response, and a
\textbf{Validator Agent} that checks grounding quality and triggers
structured recovery on failure (Figure~\ref{fig:architecture}).

Formally, given a conversation history $\mathcal{H}$ and a query $q$, the
goal is to produce an answer faithful to the evidence in $\mathcal{H}$.
For SLMs with effective context windows of 2--4K tokens, $|\mathcal{H}|$
vastly exceeds what can be reliably ingested at once. MemFlow decomposes the
problem into three decisions: \emph{what memory operation does $q$ require?}
(intent routing), \emph{what evidence should the SLM see?} (retrieval,
preprocessing, and packing), and \emph{is the answer grounded?} (validation
and escalation). This defines MemFlow as a bounded memory-control policy
rather than an open-ended agent loop. Let $\mathcal{O}$ be the finite set of
typed memory operations; each operation specifies a retrieval mode,
deterministic evidence transformation, packing budget, answer prompt, and
validation or escalation rule. The Router Agent estimates
$\pi(q) \in \mathcal{O}$, after which execution is fixed by the selected
operation. Thus, unlike model or tool routing, MemFlow does not choose among
arbitrary tools, models, or free-form plans, but selects a memory operation
whose downstream computation is deterministic. 
Crucially, the SLM drives only the intent-classification, response-generation,
and grounding-validation stages; the Memory Agent operates entirely
outside the SLM, free from SLM-driven tool selection.
The full
pipeline executes with at most four SLM calls per query, with per-stage token
statistics reported in Appendix~\ref{app:compute}.

\subsection{Router Agent and Memory Agent}
\label{sec:router_memory}

The separation between routing and execution is MemFlow's core structural
choice. Rather than asking the SLM to select tools during generation, the
Router Agent determines \emph{what} to do and the Memory Agent executes
\emph{how}. This separation avoids open-ended model-driven tool selection:
once a tag is assigned, the execution path is fixed. Our
ablation validates this design principle: disabling tag-specific retrieval
and preprocessing produces the single largest accuracy drop in the system,
costing 18.7 percentage points (Table~\ref{tab:ablation}).

\paragraph{Router Agent.}
The Router Agent classifies $q$ into one of seven typed memory operations
(\texttt{profile-injection}, \texttt{targeted-extraction},
\texttt{temporal-reasoning}, \texttt{conflict-resolution},
\texttt{broad-summarization}, \texttt{constraint-validation},
\texttt{state-tracking}) via a three-layer cascade (full prompt and tag
taxonomy in Appendix~\ref{app:router_prompt}): a rule layer fires first
on unambiguous intents; if no rule matches, a single SLM call classifies
the query; if the output fails to parse, keyword heuristics serve as a
final fallback. The cascade achieves 87.7\% routing accuracy and reduces
hard routing failures caused by malformed SLM outputs. Critically, the router has no knowledge of
downstream retrieval mechanisms or tools: its sole output is an intent
tag, ensuring that all downstream dispatch remains deterministic and
less exposed to the hallucinated tool calls that afflict open-ended
tool-selection at sub-3B scale~\cite{faghih2025tool,yao2023react}.

\paragraph{Memory Agent.}
The deterministic Memory Agent executes retrieval and preprocessing appropriate to the
action tag, then compiles the result into a token-budgeted context. It is
organized into three tiers, each corresponding to a distinct retrieval
regime: zero-retrieval for stable facts, standard retrieval for factual
lookups, and retrieval with deterministic preprocessing for complex
multi-step queries. Together, the three tiers partition all seven action
tags into exhaustive, non-overlapping execution paths~\cite{ong2024routellm,saha2024system,chen2023frugalgpt}.

\textbf{Tier 1 (Profile Lookup)} handles \texttt{profile-injection}
queries \emph{without any retrieval}. Counterintuitively, for preference
and identity facts that are stable across sessions, retrieval is
counterproductive: it risks surfacing conflicting or outdated phrasings of
the same preference. Instead, MemFlow pre-compiles these facts into a
structured user profile during ingestion and loads it directly.

\textbf{Tier 2 (Targeted Retrieval)} handles \texttt{targeted-extraction}
queries via multi-pass entity-aware retrieval. A primary hybrid
BM25+dense pass~\cite{karpukhin2020dpr,santhanam2022colbertv2} retrieves
top-$k$ chunks for the full query; secondary passes retrieve separately
for each extracted entity, handling queries that reference multiple
subjects (full hyperparameters in Appendix~\ref{app:retrieval}).

\textbf{Tier 3 (Deep Reasoning)} handles the five remaining tags, each
requiring deterministic preprocessing before the SLM sees any evidence.
The key insight is that SLMs at sub-3B scale cannot reliably perform
operations like chronological sorting, date arithmetic, or rule filtering
on raw retrieved text~\cite{liu2024lost,faghih2025tool}. MemFlow therefore completes these computations prior to any SLM inference,
as specified per-tag in Appendix~\ref{app:retrieval}: temporal queries
trigger a Chronological Sorter and Date Math Calculator; conflict queries
apply a Recency Filter that marks stale facts; summarization queries
enforce a session-diversity guarantee to prevent redundancy.

After tier execution, the Memory Agent assembles outputs into a
priority-ordered, token-budgeted context: pinned facts are never
truncated, pre-computed summaries are reduced on overflow, and raw chunks
are truncated last. The result averages 2,223 tokens overall --- a fraction of the full
history --- with allocation tables broken down by tag and benchmark in
Appendix~\ref{app:packer}.

\subsection{Answer Agent and Validator Agent}
\label{sec:answer_validator}

\paragraph{Answer Agent.}
Evidence compilation and response generation are separated into distinct
stages because the Memory Agent's output is structured specifically for
the query type, and exploiting that structure requires a matching prompt
rather than a generic one. The Answer Agent generates a response from the
packed context via a single SLM call, with a \textbf{tag-specific} system
prompt (all seven templates in Appendix~\ref{app:answer_prompts}):
temporal queries receive a chronological framing that cues the model to
narrate a timeline; profile queries receive a preference-injection
framing; extraction queries receive a grounded-recall framing. For
temporal and summarization queries, the agent may additionally emit a
bounded \texttt{TOOL:} call (e.g., \texttt{days\_between}), which is
executed deterministically and injected as a \texttt{TOOL\_RESULT} ---
the model cannot select arbitrary tools or enter open-ended loops.

\paragraph{Validator Agent.}
Even with compact, well-organized context, SLMs at sub-3B scale can
hallucinate~\cite{liu2024lost}. Prior work shows that critic-guided
verification can substantially reduce such
errors~\cite{asai2023selfrag,yan2024crag}; the Validator Agent applies
this principle without any additional training. It operates in three
stages that are intentionally ordered from cheapest to most expensive:
\textbf{(1) hard-failure detection} (deterministic, zero cost) triggers
immediate escalation on empty answers or explicit abstentions; \textbf{(2)
short-answer passthrough} (deterministic) skips grounding checks for
answers of six words or fewer; \textbf{(3) a lightweight SLM grounding
judge} (conditional) outputs a binary yes/no on whether the answer is
supported by the context (at most 8 new tokens, no chain-of-thought).
The cascade ensures that expensive LLM inference is invoked only when
necessary.

When validation fails, an \textbf{escalation loop} re-routes the query to
a heavier tier and regenerates according to a policy-driven retry schedule
described in Appendix~\ref{app:escalation_analysis}. In practice, escalation
is invoked on only 14.9\% of queries, and only 3.4\% ultimately adopt the
escalated response, indicating that most queries are handled on the first
pass while a smaller subset benefits from structured recovery.

\section{Experiments \& Results}

We evaluate \textit{MemFlow} along two axes: \emph{accuracy} on
long-horizon memory QA, and \emph{efficiency}, measured by the
answer-context tokens fed to the SLM. We conduct four experiments
targeting distinct questions: (1)~Does MemFlow outperform SLM baselines
and dedicated memory systems on established benchmarks?
(2)~Does the improvement generalize across SLM backbones?
(3)~Which pipeline components drive performance?
(4)~What does MemFlow's internal behavior look like on representative
queries? Together, these experiments validate both the system's practical
value and the design principles described in Section~\ref{sec:memflow}.

\begin{table}[t]
\centering
\caption{Main results across three long-context benchmarks.
\textbf{Answer Ctx} = tokens fed to the answer agent.
All accuracies are judged by GPT-4o-mini.
Best SLM result per column in \textbf{bold}.}
\label{tab:main_results}
\small
\resizebox{\columnwidth}{!}{%
\begin{tabular}{lccccc}
\toprule
\textbf{System} & \textbf{LongMemEval (\%)} & \textbf{LoCoMo (\%)} & \textbf{LongBench (\%)} & \textbf{AVG (\%)} & \textbf{Answer Ctx} \\
\midrule
\multicolumn{6}{l}{\textit{SLM Baselines}} \\
Direct QA (short)    & 34.2 & 24.0 & 14.6 & 21.3 & 2{,}458 \\
Direct QA (full)     & 42.2 & 33.7 & 19.8 & 29.0 & 8{,}614 \\
RAG                  & 46.0 & 45.9 & 46.7 & 46.2 & 2{,}266 \\
ReAct                & 50.2 & 49.6 & 41.2 & 46.2 & 8{,}732 \\
\midrule
\multicolumn{6}{l}{\textit{Memory Systems}} \\
Memobase             & 13.4 &  7.7 & 10.7 &  9.6 & $\sim$260\textsuperscript{\dag} \\
MemGPT               & 38.8 & 31.6 & 27.8 & 30.9 & $\sim$1{,}260\textsuperscript{\dag} \\
\midrule
\multicolumn{6}{l}{\textit{Ours}} \\
\textbf{MemFlow}     & \textbf{61.8} & \textbf{51.2} & \textbf{51.1} & \textbf{52.4} & 2{,}223 \\
\midrule
\multicolumn{6}{l}{\textit{Frontier Reference}} \\
GPT-4o (matched)     & 67.8 & 56.6 & 64.7 & 61.3 & 9{,}800$^\ast$ \\
GPT-4o (ceiling)     & 68.4 & 76.3 & 73.7 & 74.3 & 120{,}000$^\ast$ \\
\bottomrule
\end{tabular}%
}
\vspace{2pt}

{\footnotesize
All SLM Baselines and Memory Systems use Qwen3-1.7B as the backbone.
\textsuperscript{\dag}Estimated from retrieval metadata.
$^\ast$Token budget cap, not measured.
$N{=}4{,}236$ questions across three benchmarks.}
\end{table}

\subsection{Experimental Setup}

\begin{figure}[t]
    \centering
    \includegraphics[width=\columnwidth,trim=0 0 0 0, clip]{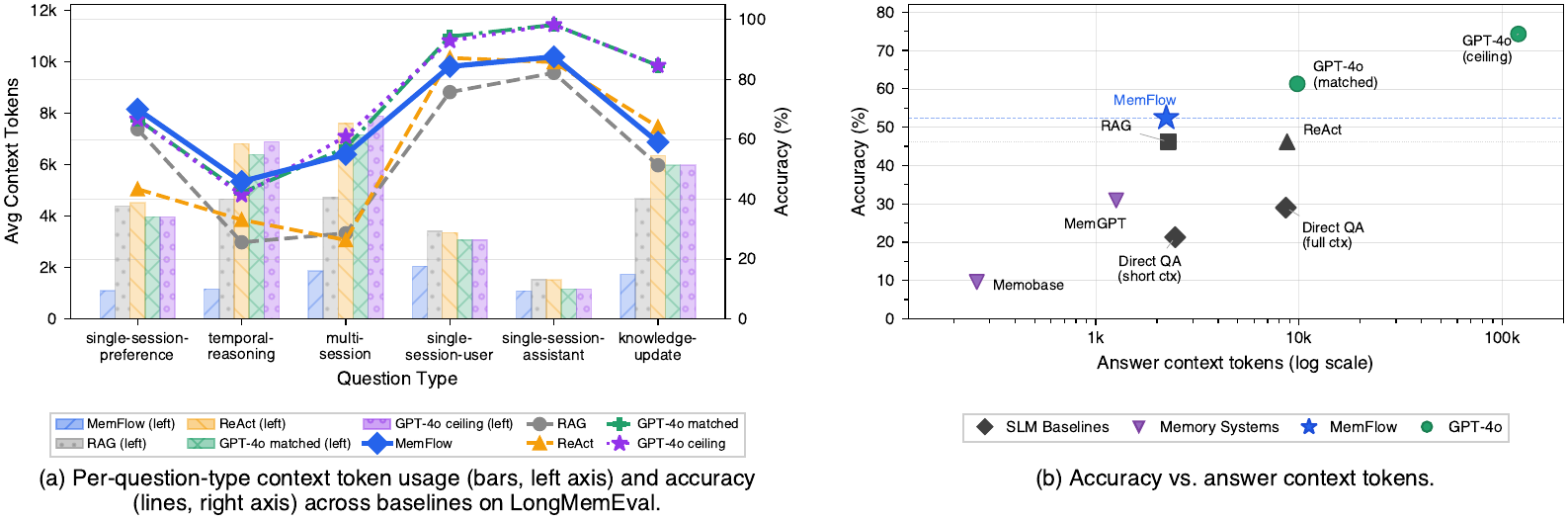}
    \caption{(a)~Per-question-type accuracy (lines, right axis) and
    mean answer-context tokens (bars, left axis) on LongMemEval.
  (b)~Accuracy vs.\ answer-context tokens across all systems (log
    scale).}
    \label{fig:accuracy_vs_context}
\vspace{2pt}
{\footnotesize }
\end{figure}

\paragraph{Benchmarks.}
We evaluate on three established long-context QA benchmarks:
\textbf{LongMemEval}~\cite{wu2024longmemeval} (500 questions across 6
question types testing long-term chat memory),
\textbf{LoCoMo}~\cite{maharana2024locomo} (1,986 questions over 600-turn
conversations spanning adversarial, multi-hop, temporal, preference, and
single-hop queries), and \textbf{LongBench}~\cite{bai2024longbench} (1,750 questions
across 9 document-level tasks). Together these yield $N{=}4{,}236$
evaluation items covering both session-based episodic memory and
single-document comprehension.

\paragraph{Model and baselines.}
All SLM systems use \textbf{Qwen3-1.7B}~\cite{qwen2025qwen3} as the frozen
backbone unless otherwise noted. We compare against four SLM baselines:
\textbf{Direct QA (short/full)}, with raw conversation truncated to
match MemFlow's answer-context and full-pipeline token budgets
respectively; \textbf{RAG}~\cite{lewis2020rag}, hybrid BM25+dense
retrieval with top-$k$ chunks (2,266 avg tokens); and
\textbf{ReAct}~\cite{yao2023react}, multi-step reasoning with access to
the same retrieval and deterministic tools (8,732 avg tokens). We further compare against two dedicated
memory systems using the same backbone:
\textbf{Memobase}~\cite{chhikara2025mem0} (user-profile memory with BM25
retrieval) and \textbf{MemGPT}~\cite{packer2023memgpt} (working+archival
memory with model-driven retrieval).
\textbf{GPT-4o}~\cite{openai2024gpt4o} results are reported at matched
(9.8k) and full (120k) token budgets as frontier references.
All SLM baselines use the same benchmark inputs and GPT-4o-mini judge; RAG and
ReAct use the same hybrid BM25+dense retrieval backend as MemFlow. For external
memory systems, context sizes are estimated from retrieval metadata where
direct token accounting is unavailable; these comparisons should therefore be
interpreted as system-level references rather than exact efficiency-matched
baselines. Full implementation details are in Appendix~\ref{app:baselines}.

\paragraph{Evaluation.}
All systems are scored by a GPT-4o-mini
judge~\cite{openai2024gpt4omini} that compares predicted answers against
gold references with a binary correct/incorrect judgment. We use deterministic
decoding over the full $N{=}4{,}236$ question pool and report accuracy (\%)
as the primary metric; exact match and F1 are
not reported as they systematically underestimate performance on
open-ended memory QA where semantically correct answers differ lexically
from gold references~\cite{wu2024longmemeval}. To check judge reliability,
we compare GPT-4o-mini and GPT-4o on 300 held-out predictions, observing
85.7\% agreement; disagreements are mostly cases where GPT-4o-mini is
stricter (Appendix~\ref{app:judge}). We also manually annotate a held-out sample of 300
predictions stratified across benchmarks and systems; GPT-4o-mini agrees with
human labels on 89.6\% of cases, with most disagreements involving
borderline semantic-equivalence judgments (Appendix~\ref{app:judge}). We
therefore treat LLM-judge accuracy as a semantic correctness estimate rather
than an exact lexical benchmark score.


\subsection{Main Results}
\label{sec:main_results}

Table~\ref{tab:main_results} summarizes average accuracy across the three
benchmarks. \textbf{MemFlow achieves 52.4\% overall}, the highest among
all SLM-based systems, representing a $+$23.4\,pp lift over Direct QA
(full) and $+$6.2\,pp over both RAG and ReAct. RAG and ReAct happen to
tie at 46.2\% on average (a coincidence of aggregation, as their
per-type profiles differ considerably), yet ReAct uses $3.9\times$ as many tokens (8,732 vs.\ 2,266), indicating
that open-ended reasoning loops provide no net benefit over simple retrieval
for SLM-scale memory tasks.

Among dedicated memory systems, MemGPT reaches 30.9\% and Memobase only
9.6\%---both substantially below MemFlow despite sharing the same
backbone. The performance gap is architectural: Memobase's profile
compression discards episodic recall ($\sim$260 context tokens), while
MemGPT's unstructured retrieval ($\sim$1,260 tokens) lacks the
intent-aware routing needed to surface relevant evidence.

\textbf{Context and pipeline cost.}
Figure~\ref{fig:accuracy_vs_context}(b) plots accuracy against
answer-context tokens, i.e., the evidence budget seen by the final answer
model. By this measure, MemFlow reaches the highest SLM accuracy at an
answer context comparable to RAG (2,223 vs.\ 2,266 tokens) and far smaller
than ReAct (2,223 vs.\ 8,732 tokens). This is not the same as total inference
cost: the full MemFlow pipeline averages 10,167 prompt tokens per query
across the router, answer agent, validator, and conditional escalation
(Appendix~\ref{app:compute}), slightly above ReAct's 8,732-token reasoning
loop. Thus MemFlow should be interpreted as trading a modest amount of
orchestration overhead for a more reliable and compact final reasoning
context: it improves average accuracy by $+6.2$\,pp over ReAct while reducing
the context seen by the final answer stage by $3.9\times$. We report both
views because they measure different costs: full pipeline tokens capture
system-level inference overhead, while answer-context length measures the
reasoning burden placed on the small model. Figure~\ref{fig:accuracy_vs_context}(a)
further shows that the answer-context reduction holds across all six
LongMemEval question types. The gap to budget-matched GPT-4o is 8.9\,pp
(52.4\% vs.\ 61.3\%), representing the residual cost of compressing a long
conversation into a compact packed context rather than attending to the full
history with a substantially larger model.

\textbf{Per-question-type breakdown.}
MemFlow matches or exceeds budget-matched GPT-4o on
\texttt{single-session-preference} ($+$3.3\,pp) and
\texttt{temporal-reasoning} ($+$3.8\,pp) on LongMemEval 
(Figure~\ref{fig:accuracy_vs_context}(a)), and dominates
adversarial queries on LoCoMo ($+$8.7\,pp over ReAct), where the
Validator Agent's grounding check rejects hallucinated answers before
they are returned. Remaining gaps to GPT-4o are concentrated on
\texttt{knowledge-update} ($-$25.6\,pp), where cross-session conflict
resolution exceeds the packer's compression fidelity, and on queries routed to
the \texttt{conflict-resolution} tier ($-$48.9\,pp relative to GPT-4o on those items),
driven partly by router over-prediction on that tag. Full per-type results are in
Appendix~\ref{app:per_type}.

\subsection{Generalization Across SLM Backbones}

\begin{table}[t]
\centering
\caption{MemFlow generalizes across SLM backbones. Each model is
evaluated in Direct QA mode and with MemFlow
scaffolding. $\Delta$ = absolute improvement from MemFlow.
All accuracies judged by GPT-4o-mini.}
\label{tab:slm_comparison}
\small
\resizebox{\columnwidth}{!}{%
\begin{tabular}{llccccc}
\toprule
\textbf{Backbone} & \textbf{Mode} & \textbf{LongMemEval (\%)} & \textbf{LoCoMo (\%)} & \textbf{LongBench (\%)} & \textbf{Overall (\%)} & $\boldsymbol{\Delta}$ \\
\midrule
\multirow{2}{*}{Qwen3-1.7B}
  & Direct  & 42.2 & 33.7 & 19.8 & 29.0 & \\
  & MemFlow & \textbf{61.8} & \textbf{51.2} & \textbf{51.1} & \textbf{52.4} & \textbf{+23.4} \\
\midrule
\multirow{2}{*}{SmolLM2-1.7B}
  & Direct  & 25.0 & 18.6 & 19.3 & 19.6 & \\
  & MemFlow & \textbf{37.2} & \textbf{47.9} & \textbf{39.0} & \textbf{43.0} & \textbf{+23.4} \\
\midrule
\multirow{2}{*}{Qwen3-0.6B}
  & Direct  & 26.6 & 23.8 & 34.0 & 28.3 & \\
  & MemFlow & \textbf{50.6} & \textbf{41.6} & \textbf{38.5} & \textbf{41.4} & \textbf{+13.1} \\
\midrule
\multirow{2}{*}{LLaMA-3.2-1B}
  & Direct  & 20.6 & 21.4 & 16.9 & 19.4 & \\
  & MemFlow & \textbf{38.2} & \textbf{45.5} & \textbf{32.1} & \textbf{39.1} & \textbf{+19.7} \\
\midrule
\multirow{2}{*}{Gemma-3-1B}
  & Direct  & 24.6 & 21.6 & 25.0 & 23.4 & \\
  & MemFlow & \textbf{28.6} & \textbf{27.9} & \textbf{31.4} & \textbf{29.4} & \textbf{+6.0} \\
\bottomrule
\end{tabular}%
}
\vspace{2pt}

{\footnotesize
Direct QA uses the available direct-prompt setting for each backbone.
MemFlow answer context averages $\sim$2{,}200 tokens across all backbones.
Models ordered by Overall MemFlow accuracy.}
\end{table}

Table~\ref{tab:slm_comparison} evaluates MemFlow with four additional SLM
backbones: Qwen3-0.6B~\cite{qwen2025qwen3}, SmolLM2-1.7B~\cite{allal2025smollm2},
LLaMA-3.2-1B~\cite{meta2024llama32}, and Gemma-3-1B~\cite{gemma2025gemma3}.
MemFlow improves every backbone without exception, with lifts ranging from
$+$6.0\,pp (Gemma-3-1B) to $+$23.4\,pp (Qwen3-1.7B and SmolLM2-1.7B). 
The largest gains occur on models that produce near-degenerate outputs
under direct prompting: SmolLM2-1.7B and LLaMA-3.2-1B both score under
20\% on direct QA, with manual inspection revealing near-empty or
format-broken responses, yet reach 43.0\% and 39.1\% under MemFlow.
This confirms that MemFlow's structured context injection restores
coherent output formatting regardless of backbone capacity.
Gemma-3-1B receives the smallest lift ($+$6.0\,pp), suggesting it handles
the structured prompt format less effectively than other architectures.
Notably, all backbones benefit consistently across all three benchmarks,
confirming that MemFlow's gains are not specific to any one model or
evaluation domain.
All of these gains are obtained without model fine-tuning, reinforcement
learning, or benchmark-specific training. The same orchestration pipeline is
applied across LongMemEval, LoCoMo, and LongBench; we therefore interpret the
backbone results as evidence that the improvements come from structured
memory operations rather than adaptation to a single model family.



\subsection{Ablation study}
\begin{table}[t]
\centering
\caption{Ablation study. Each row removes one component from the full
MemFlow pipeline. $\Delta$ = change in overall accuracy relative to
full MemFlow. All accuracies judged by GPT-4o-mini.}
\label{tab:ablation}
\small
\resizebox{\columnwidth}{!}{%
\begin{tabular}{lccccc}
\toprule
\textbf{Configuration} & \textbf{LongMemEval (\%)} & \textbf{LoCoMo (\%)} & \textbf{LongBench (\%)} & \textbf{Overall (\%)} & $\boldsymbol{\Delta}$ \\
\midrule
MemFlow (full)            & 61.8 & 51.2 & 51.1 & \textbf{52.4} & --- \\
\midrule
-- Retrieval Strategy     & 45.6 & 33.9 & 30.1 & 33.7 & $-$18.7 \\
-- Uniform RAG            & 60.4 & 42.8 & 37.9 & 42.9 & $-$9.5 \\
-- Router                 & 56.0 & 45.3 & 39.2 & 44.0 & $-$8.4 \\
-- Tools                  & 60.8 & 45.1 & 37.9 & 44.0 & $-$8.4 \\
-- Escalation \& Validator & 58.2 & 44.3 & 41.2 & 44.7 & $-$7.7 \\
-- Packer                 & 58.0 & 43.0 & 43.1 & 44.8 & $-$7.6 \\
\bottomrule
\end{tabular}%
}
\vspace{2pt}

{\footnotesize
Rows ordered by impact ($|\Delta|$). All ablations use the same Qwen3-1.7B
backbone, evaluation set, judge, and approximately matched answer-context
budget.}
\end{table}

Table~\ref{tab:ablation} isolates the contribution of each component under the same Qwen3-1.7B backbone, evaluation set, judge, and approximately matched answer-context budget. \textbf{-- Retrieval Strategy} disables tag-specific retrieval and preprocessing while retaining the rest of the pipeline; \textbf{-- Uniform RAG} replaces the tiered Memory Agent with flat hybrid BM25+dense retrieval; \textbf{-- Router} bypasses intent classification and uses a fixed fallback memory path; \textbf{-- Tools} removes deterministic preprocessing and bounded tool calls such as date arithmetic and counting; \textbf{-- Escalation \& Validator} returns the first answer without grounding checks or heavier-tier retry; and \textbf{-- Packer} replaces priority-aware context compilation with simple retrieved-chunk concatenation.

\textbf{Retrieval strategy selection} is the single most critical component ($-$18.7\,pp): disabling tag-specific retrieval and preprocessing causes the largest drop. \textbf{Router, Uniform RAG, and Tools} form a cluster at $-$8.4 to $-$9.5\,pp, confirming that routing quality, retrieval specialization, and deterministic preprocessing are tightly coupled; the explicit \textbf{-- Uniform RAG} row isolates the flat-retrieval replacement ($-$9.5\,pp). \textbf{Escalation/Validator and Packer} each contribute a meaningful margin ($-$7.6 to $-$7.7\,pp), validating that grounding verification and priority-ordered context compilation are both necessary. Token costs remain stable across ablations (within $\pm$5\%), suggesting that MemFlow's gains derive from the \emph{quality} of retrieved context rather than the quantity of computation. Per-benchmark breakdowns are in Appendix~\ref{app:ablation_bench}.

\subsection{Qualitative analysis}

\begin{figure*}[t]
    \centering
    \includegraphics[width=\textwidth]{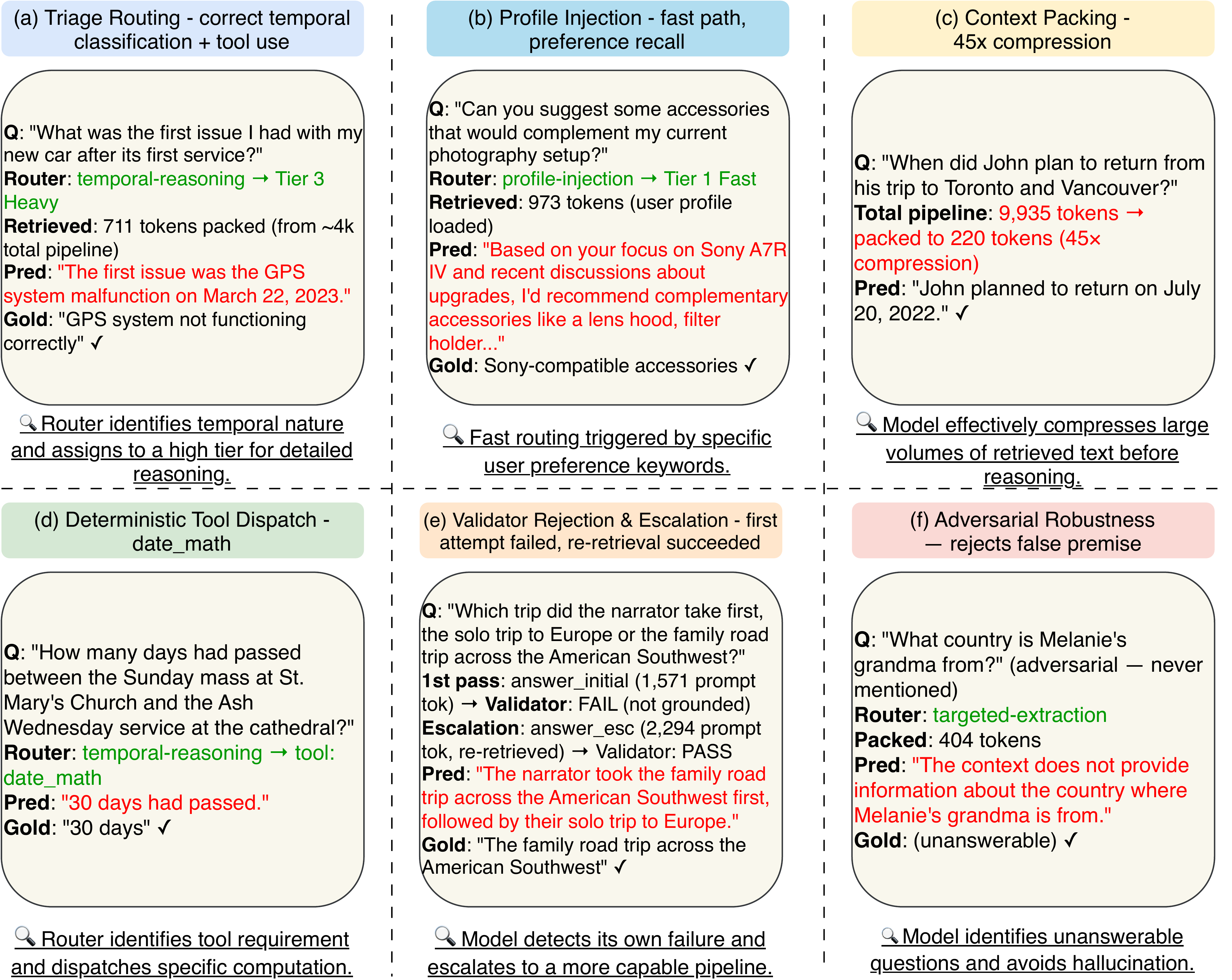}
    \caption{Snippets of MemFlow's internal states and actions on
    representative queries. Each panel shows a distinct pipeline behavior:
    (a)~triage routing, (b)~profile injection, (c)~context packing,
    (d)~deterministic tool dispatch, (e)~validator rejection and
    escalation, (f)~adversarial robustness. All examples are drawn from
    actual predictions on the LongMemEval, LoCoMo, and LongBench
    benchmarks.}
    
    \label{fig:snippets}
\end{figure*}

Figure~\ref{fig:snippets} illustrates MemFlow's internal behavior on six
representative examples. Panel~(a) shows the router correctly dispatching
a temporal query to a Tier~3 handler with tool access, producing the
correct answer from only 711 packed tokens. Panel~(c) demonstrates
$45\times$ context compression (9,935$\to$220 tokens) with no information
loss --- the packed context contains exactly the evidence needed to answer
the query. Panel~(e) shows the validator--escalation loop recovering from
an initially ungrounded answer: the first pass (1,571 prompt tokens) fails
validation, triggering re-retrieval with a broader strategy (2,294 tokens)
that produces the correct answer. Panel~(f) confirms adversarial
robustness: when queried about information absent from the conversation,
MemFlow correctly abstains rather than hallucinating --- a failure mode
that afflicts both Direct QA and ReAct on the same questions.

\section{Conclusion, Limitations, and Future Work}

We presented \textit{MemFlow}, a training-free memory orchestration framework for long-horizon SLM agents. Its core finding is that a substantial portion of SLM memory failure can be mitigated by matching each query to an appropriate memory operation before evidence is retrieved, transformed, and packed. Rather than introducing a new retriever, compressor, or verifier in isolation, \textit{MemFlow} shows that bounded orchestration---structured intent routing, deterministic evidence compilation, and grounding validation---can help frozen SLMs use limited context more effectively and narrow the gap to frontier models in targeted long-horizon memory settings. Although motivated by personal and conversational memory, the same operation-routing view extends to document QA and assistant settings where small models answer from large, heterogeneous context.
Several limitations remain: the full pipeline has nontrivial orchestration overhead dominated by the router prompt, single-intent routing can miss compound queries, evolving cross-session facts remain difficult, and LLM-judge accuracy is a semantic rather than exact lexical score. A promising future direction is to reduce router overhead and incorporate
learned decision policies for validation, escalation, and tool dispatch, using
answer correctness as a reward signal while preserving MemFlow's bounded
execution paths and compact final-answer contexts.



\clearpage
\bibliographystyle{plain}
\bibliography{references}

\newpage
\appendix

\section{System Implementation Details}
\label{app:system}

\subsection{Computational Setup}
\label{app:compute}

All MemFlow experiments are conducted on \textbf{Google Colab} using NVIDIA \textbf{L4}, \textbf{A100}, and \textbf{H100} GPUs
(whichever was available at runtime; results are hardware-independent since no training is performed).
The frozen \textbf{Qwen3-1.7B}~\cite{qwen2025qwen3} backbone is loaded via the \texttt{transformers} inference API with
\texttt{bfloat16} precision and served on GPU.
The dense retrieval encoder (\texttt{BAAI/bge-small-en-v1.5}) runs on \textbf{CPU} only;
GPU inference for the encoder was disabled because LoCoMo's 35-session conversations
exhaust GPU memory during FAISS index construction on smaller GPU tiers.
No training is performed: all model weights are frozen and identical across all system variants and GPU configurations.

\paragraph{Token cost breakdown.}
Table~\ref{tab:token_breakdown} decomposes the per-query token usage across pipeline stages,
computed from the full $N{=}4{,}236$ evaluation set.

\begin{table}[h]
\centering
\small
\caption{Per-stage token usage, MemFlow main experiment ($N=4{,}236$, Qwen3 tokenizer).
$\dagger$~2.0\% of queries short-circuit before routing/answering/validation; the remaining router reduction comes from the rule layer.
$\ddagger$~Of the 14.9\% that trigger escalation, 3.4\% receive the escalated response as the final answer.
$*$~Weighted contribution to mean pipeline cost after multiplying conditional mean tokens by invocation rate.}
\label{tab:token_breakdown}
\begin{tabular}{lrrrr}
\toprule
Stage & Mean tokens & Median & Weighted \% & Invocation rate \\
\midrule
Router Agent (prompt)     & 6{,}954  & 6{,}953 & 58.0\%\textsuperscript{$*$}  & 84.8\%\textsuperscript{$\dagger$} \\
Answer Agent (prompt)     & 2{,}550  & 1{,}468  & 24.6\%\textsuperscript{$*$}  & 98.0\%\textsuperscript{$\dagger$}  \\
Validator (prompt)        & 1{,}087  & 1{,}075  & 10.5\%\textsuperscript{$*$}  & 98.0\%\textsuperscript{$\dagger$} \\
Escalation answer (prompt)& 2{,}289  & 2{,}260  & 3.4\%\textsuperscript{$*$}  & 14.9\%\textsuperscript{$\ddagger$} \\
Fixed wrappers/tool results & 314 & --- & 3.1\% & 100\% \\
\midrule
\textbf{Total pipeline}   & \textbf{10{,}167} & 8{,}793 & --- & --- \\
\textbf{Packed context (answer input)} & \textbf{2{,}223} & 874 & --- & --- \\
\bottomrule
\end{tabular}
\end{table}

The Router Agent accounts for the largest weighted share of token usage (58.0\% of the pipeline mean),
driven by its fixed system prompt that encodes the full seven-tag taxonomy.
In deployments that serve repeated queries with the same router prompt, this static prefix
can be reused through prefix/KV caching, reducing prefill overhead even though the token
accounting in Table~\ref{tab:token_breakdown} reports uncached prompt tokens.
The Answer Agent's mean prompt of 2,550 tokens exceeds its median of 1,468 tokens, indicating
that most queries are handled with relatively compact packed contexts while a subset of
complex queries receives substantially larger inputs.
The Validator is invoked on 98.0\% of queries at an average of 1,087 tokens per call (10.5\%
weighted contribution).
The Escalation answer stage fires on 14.9\% of queries; its 3.4\% weighted contribution to the
mean pipeline cost is computed as $2{,}289 \times 14.9\% / 10{,}167$.
The remaining 314-token average covers fixed chat-template wrappers, query fields,
and tool-result prompts not assigned to a single agent row.
Of those 14.9\% escalated queries, 3.4\% of \emph{all} queries (146/4{,}236) ultimately
receive a different escalated response as the final answer (\texttt{is\_escalated = True});
the remaining 11.5\% trigger validation-driven retry but retain the original answer or a fallback.

\paragraph{Per-benchmark packed context.}
The 2{,}223-token average masks substantial variation across benchmarks, reflecting the
adaptive nature of the packer: LongMemEval (session-based episodic queries) averages
1{,}543 tokens; LoCoMo (peer-conversation queries) averages 469 tokens; LongBench
(document-level QA and summarization) averages 4{,}408 tokens, approaching the 6{,}000-token
Tier-2 ceiling for broad-coverage document tasks.

\subsection{Retrieval Configuration}
\label{app:retrieval}

The retrieval layer uses \textbf{hybrid BM25+dense} retrieval with Reciprocal Rank Fusion (RRF).
Table~\ref{tab:retrieval_cfg} lists all hyperparameters.

\begin{table}[h]
\centering
\small
\caption{Retrieval hyperparameters.}
\label{tab:retrieval_cfg}
\begin{tabular}{lll}
\toprule
Parameter & Value & Notes \\
\midrule
Strategy          & \texttt{hybrid}               & BM25 + dense, fused by RRF \\
Dense encoder     & \texttt{BAAI/bge-small-en-v1.5} & 384-dim, CPU inference \\
BM25 $k_1$        & 1.5                           & Okapi BM25 saturation \\
BM25 $b$          & 0.75                          & Length normalization \\
RRF $k$           & 60                            & Standard RRF constant \\
Base top-$k$      & 8                             & Candidates per hybrid pass \\
Tier-2 top-$k$    & 20                            & Targeted extraction \\
Tier-2 doc top-$k$& 40                            & LongBench document tasks \\
Tier-3 top-$k$    & 20                            & Heavy reasoning passes \\
Tier-3 broad top-$k$ & 80                         & Broad-summarization tag \\
Map chunk size    & 5                             & Turns per map-reduce shard \\
\bottomrule
\end{tabular}
\end{table}

For Tier-2 \textit{targeted-extraction} queries, the handler performs \textbf{multi-pass entity-aware
retrieval}: a primary hybrid pass retrieves the top-$k$ chunks for the full query; secondary passes
retrieve separately for each extracted entity or noun phrase; results are unioned, deduplicated by
chunk ID, and ranked by score.
For Tier-3 \textit{temporal-reasoning} queries, \textbf{dual-anchor retrieval} extracts event
references from the query and retrieves separately for each anchor before merging.
Entity and noun-phrase anchors are extracted with a deterministic lightweight parser:
capitalized name spans, quoted strings, numeric/date expressions, and noun phrases around
possessive or relational markers are retained after stopword filtering. Event anchors for
temporal queries are the two highest-scoring verb-phrase or date-bearing spans that overlap
with question terms. No external trained parser is used.
For the remaining Tier-3 tags, \textit{conflict-resolution} sorts candidate evidence by
timestamp and marks stale facts, \textit{broad-summarization} applies session-diverse
map-reduce aggregation, \textit{constraint-validation} filters for modal and rule language
(``always'', ``never'', ``must'', ``allowed''), and \textit{state-tracking} keeps chronological
state-change candidates before packing.

\subsection{Context Packer: Detailed Budget Allocation}
\label{app:packer}

The Context Packer enforces a strict global SLM ceiling and allocates it across three priority
slots (ordered below by packer priority, not by retrieval tier number):
(1)~\textbf{Tier-1} slots (pinned facts, user profile, behavioral constraints; never truncated),
(2)~\textbf{Tier-3} slots (pre-computed summaries and aggregated facts; reduced on overflow), and
(3)~\textbf{Tier-2} slots (raw episodic chunks from retrieval; truncated first on overflow).
Unused budget from higher-priority tiers is dynamically reallocated to lower tiers.
The resulting packed context averages $\sim$2,200 tokens across all benchmarks
(see Table~\ref{tab:bench_tokens} for per-benchmark breakdowns).

\paragraph{Per-tag Tier-2 budgets.}
Different \texttt{action\_tag} values impose different Tier-2 chunk budgets, reflecting the
information density required for each query type:

\begin{table}[h]
\centering
\small
\caption{Per-tag Tier-2 chunk budgets and per-chunk word caps.
Dashes indicate no per-chunk word cap is applied.}
\label{tab:tag_budgets}
\begin{tabular}{lrr}
\toprule
Action tag & Tier-2 budget (tokens) & Per-chunk word cap \\
\midrule
\texttt{profile-injection}       & 0 & N/A \\
\texttt{targeted-extraction}     & 6{,}000 & 300 words \\
\texttt{temporal-reasoning}      & 4{,}400 & --- \\
\texttt{conflict-resolution}     & 6{,}000 & --- \\
\texttt{broad-summarization}     & 8{,}000 & --- \\
\texttt{constraint-validation}   & 6{,}000 & 200 words \\
\texttt{state-tracking}          & 6{,}000 & 150 words \\
\bottomrule
\end{tabular}
\end{table}

\paragraph{Packer enhancements.}
Beyond budget enforcement, the packer applies four post-retrieval enhancements:
(1) \textbf{Relevance sort} --- chunks are ordered by reranker score before packing;
(2) \textbf{Sentence-level extraction} --- for precision-sensitive tags, the packer keeps
only the top-2 most relevant sentences per chunk rather than the full chunk;
(3) \textbf{Jaccard deduplication} --- near-duplicate chunks (Jaccard similarity $\geq 0.85$)
are dropped; and
(4) \textbf{Dynamic budget reallocation} --- unused budget from Tier-1 or Tier-3 sections
is reallocated to the Tier-2 episodic section.
\texttt{targeted-extraction}, \texttt{constraint-validation}, and \texttt{state-tracking}
are treated as precision-sensitive because the answer usually depends on a specific fact,
rule, or state transition rather than a broad summary. Sentence relevance is scored by
normalizing the sum of dense similarity and token-overlap with the query. The top-2
sentence cap and Jaccard threshold of 0.85 are tuned implementation parameters selected
on development runs to reduce duplicate evidence under the SLM budget; no test-set
labels are used to choose them.




\section{Complete Prompt Listings}
\label{app:prompts}

\subsection{Router Agent System Prompt}
\label{app:router_prompt}

The Router Agent receives the user query and outputs a structured JSON routing decision.
The full system prompt encodes the seven-tag taxonomy, disambiguation rules, and few-shot
JSON examples.
Below we reproduce the tag definitions and selected disambiguation rules; the complete prompt
(with all examples) is released with the code.

\paragraph{Tag taxonomy.}
\begin{itemize}[leftmargin=1.5em]
\item \textbf{\texttt{profile-injection}} --- the user asks for a draft, recommendation, or action
  that should reflect their personal style, tone, or stated preferences.
  No retrieval required; the answer comes from a stored user profile.

\item \textbf{\texttt{targeted-extraction}} --- the user wants to recall one specific fact,
  detail, item, or piece of content from a single past conversation.
  The fact is stable and not subject to change.

\item \textbf{\texttt{temporal-reasoning}} --- the user asks about the order or timing of past
  events: which came first, how long between two events, when something happened relative
  to something else, or how much time elapsed.
  Requires chronological sorting and/or date arithmetic.

\item \textbf{\texttt{conflict-resolution}} --- the user asks for a value that may have changed
  over time and wants the most recent value (e.g.\ current job, latest address, updated plan).
  Must retrieve across sessions and select the newest version.

\item \textbf{\texttt{broad-summarization}} --- the user asks for a count, list, frequency, or
  synthesis of something mentioned across multiple past conversations.
  Requires scanning and combining facts from distinct sessions.

\item \textbf{\texttt{constraint-validation}} --- the user asks about a behavioral rule, policy,
  or constraint: what is always/never allowed, what rule applies to a situation.

\item \textbf{\texttt{state-tracking}} --- the user asks how a fact, preference, or situation
  \emph{evolved} or \emph{changed} over time: the full progression, not just the current state.
\end{itemize}

\paragraph{Output schema.}
The router outputs a constrained JSON object:
\begin{verbatim}
{
  "requires_rag": boolean,
  "requires_reasoning": boolean,
  "action_tag": string  // one of the seven tags above
}
\end{verbatim}
If the SLM output cannot be parsed as valid JSON, a regex rescue extracts the tag from
free-form text; a keyword heuristic serves as the final fallback.
This three-tier cascade (rules $\to$ SLM $\to$ heuristic) achieves 87.7\% overall routing
accuracy (\S{}\ref{app:router_accuracy}).

\paragraph{Selected disambiguation rules (abridged).}
The prompt encodes explicit disambiguation rules to resolve frequent boundary cases.
A representative selection:

\begin{itemize}[leftmargin=1.5em]
\item \textit{``How many days/weeks passed between X and Y?''} $\to$ \texttt{temporal-reasoning}
  (elapsed-time measurement, not event counting).

\item \textit{``How long was I in Japan?''} $\to$ \texttt{targeted-extraction}
  (duration of a single event = one stable recalled fact, not temporal comparison).

\item \textit{``What is my current job?''} $\to$ \texttt{conflict-resolution}
  (one tracked thing whose value may have changed; need most recent).

\item \textit{``How many magazine subscriptions do I have?''} $\to$ \texttt{broad-summarization}
  (many distinct items each acquired in a different conversation; requires broad scan).

\item \textit{``What should I do when a customer complains?''} $\to$ \texttt{constraint-validation}
  (procedural rule for a conditional situation, not open-ended advice).

\item \textit{``Any tips for keeping my kitchen clean?''} $\to$ \texttt{profile-injection}
  (advice request without an explicit condition; answer depends on user preferences).
\end{itemize}

\paragraph{Programmatic rule layer.}
Before the SLM router is called, deterministic rules handle unambiguous lexical patterns:
elapsed-time questions with ``how many days/weeks/months between'' map to
\texttt{temporal-reasoning}; ``current/latest/most recent'' value questions map to
\texttt{conflict-resolution}; explicit ``always/never/allowed/must'' policy questions map to
\texttt{constraint-validation}; ``how has X changed/evolved'' maps to \texttt{state-tracking};
and personalized drafting or recommendation requests with user-preference cues map to
\texttt{profile-injection}. If multiple rules match, the query is sent to the SLM router rather
than resolved by the rule layer.

\subsection{Per-Tag Answer Agent System Prompts}
\label{app:answer_prompts}

The Answer Agent receives a tag-specific system prompt that frames the reasoning task.
We provide the full prompt text for each of the seven primary tags below.

\paragraph{\texttt{profile-injection}.}
\begin{quote}\small
\textit{``You are a personalized assistant.
First, identify the user's stated preferences, past choices, or personal details
from the conversation history in the context.
Then give a tailored response that reflects those specific preferences.
If the question asks what the user would prefer or what kind of response they want,
describe their preferences explicitly (e.g.\ `Based on your past conversations, you prefer X').
If there is only partial information, still give your best tailored response and
cite the specific preference or fact you are drawing on.
Prefer a personalized answer whenever the context contains any relevant preference.
Keep the response concise (1--3 sentences).
[shared grounding instruction]''}
\end{quote}

\paragraph{\texttt{targeted-extraction}.}
\begin{quote}\small
\textit{``You are answering a targeted memory-recall question.
Use only the retrieved context to identify the specific fact requested by the user.
If multiple context snippets mention the same entity, choose the snippet that most directly
answers the question and quote or paraphrase only that fact.
Do not add background details that are not needed for the answer.
[shared grounding instruction]''}
\end{quote}

\paragraph{\texttt{temporal-reasoning} (tool mode, elapsed-time queries).}
This prompt enforces a three-step protocol when date arithmetic is required:
\begin{quote}\small
\textit{``STEP 1 --- IDENTIFY BOTH DATES: Scan the context and write down the date for each
event mentioned in the question.
State them explicitly: `Event A date: YYYY-MM-DD (from [quote from context])'.\quad
STEP 2 --- CALL THE TOOL: After identifying both dates, call exactly ONE tool:
\texttt{TOOL: days\_between | YYYY-MM-DD | YYYY-MM-DD}
(or \texttt{weeks\_between} / \texttt{months\_between}).\quad
STEP 3 --- FINAL ANSWER: After the TOOL\_RESULT line appears, give your answer using the
computed number. [Additional date extraction rules and shared grounding instruction]''}
\end{quote}

\paragraph{\texttt{conflict-resolution}.}
\begin{quote}\small
\textit{``You are answering with latest-only conflict-resolved memory facts.
The context is sorted chronologically --- the most recent entries appear last.
Always use the most recent value when facts conflict.
[shared grounding instruction]''}
\end{quote}

\paragraph{\texttt{broad-summarization} (tool mode, counting queries).}
\begin{quote}\small
\textit{``You are answering a question that may require counting or aggregating facts
across multiple past conversations. The context contains map-reduced cross-session summaries.
You have access to a counting tool. If the question asks how many times a specific
thing appears, call: \texttt{TOOL: count\_occurrences | keyword}.
Call only one tool per response. After receiving TOOL\_RESULT, give your final answer.
[shared grounding instruction]''}
\end{quote}

\paragraph{\texttt{constraint-validation}.}
\begin{quote}\small
\textit{``You are answering a question about behavioral rules, policies, or constraints.
List all applicable Always/Never rules that answer the question and cite the exact rule text.
If multiple rules apply, list each one.
If no constraint is found in context, say so explicitly.
[shared grounding instruction]''}
\end{quote}

\paragraph{\texttt{state-tracking}.}
\begin{quote}\small
\textit{``You are answering a question about how something changed or evolved over time.
The context is sorted chronologically --- earlier sessions appear first.
Present the full progression: describe how the fact or situation changed at each stage.
Be specific about what changed and when (by session or date if available).
Do not just give the final state --- show the journey.
[shared grounding instruction]''}
\end{quote}

\paragraph{Shared grounding instruction (appended to all prompts).}
\begin{quote}\small
\textit{``You are a helpful assistant. Answer using ONLY the information in the provided context.
If you have partial or relevant information, always attempt your best answer ---
do not default to ESCALATE\_REQUIRED out of uncertainty.
Output the exact literal string ESCALATE\_REQUIRED ONLY when the context contains
absolutely no information relevant to the question.
Keep the answer concise.''}
\end{quote}

\paragraph{LoCoMo peer-conversation override.}
For LoCoMo items, the context is a dialogue between two named speakers (e.g.\ Alice and Bob)
discussing third-party people (e.g.\ Caroline, Melanie).
A specialized prefix instructs the model to search every turn for mentions of the third party
and extract facts embedded in narrative speech (e.g.\ \textit{``I talked to X and she mentioned Y''}),
and to avoid premature ESCALATE\_REQUIRED output.

\subsection{Validator System Prompt and Cascade Logic}
\label{app:validator_prompt}

The Validator uses the following system prompt for its LLM grounding check (Qwen3-1.7B,
\texttt{max\_new\_tokens=8}, \texttt{temperature=0.0}):

\begin{quote}\small
\textit{``You are a grounding verifier for a memory retrieval system.
Given a question, retrieved context, and a candidate answer, decide whether
the candidate answer is supported by --- or can reasonably be inferred from ---
the retrieved context.
Reply with exactly one word: `yes' if supported, `no' if not supported.''}
\end{quote}

The user message template is:
\begin{verbatim}
Question: {question}

Retrieved context:
{ctx_snippet}   [truncated to 6,000 characters]

Candidate answer: {answer}

Is the candidate answer supported by the retrieved context? Reply yes or no.
\end{verbatim}

\paragraph{Three-stage cascade.}
The validator applies the following cascade before invoking the LLM:
\begin{enumerate}[leftmargin=1.5em]
  \item \textbf{Hard-failure detection.}
    Empty answers, the exact string \texttt{ESCALATE\_REQUIRED} (or variants matched
    by a regex), and ``not found'' patterns immediately trigger escalation.
    No LLM call is made.

  \item \textbf{Short-answer passthrough.}
    Answers of $\leq$6 words, purely numeric answers (e.g.\ \textit{``5''}, \textit{``21 days''}),
    and boolean answers (\textit{``yes''}/\textit{``no''}) bypass the grounding check entirely.
    Single-number or short factual extractions are almost always correct when they appear
    in the answer agent's output.

  \item \textbf{LLM grounding check.}
    For all remaining answers, the validator calls Qwen3-1.7B with the prompt above.
    If the response cannot be parsed as yes/no, the validator falls back to a
    token-overlap heuristic ($\tau_{\text{ground}} = 0.07$).
\end{enumerate}

\section{Benchmark and Evaluation Details}
\label{app:benchmarks}

\subsection{Benchmark Composition}
\label{app:bench_composition}

Table~\ref{tab:bench_stats} summarizes the three benchmarks used in our evaluation.

\begin{table}[h]
\centering
\small
\caption{Benchmark statistics. ``Session depth'' refers to the number of conversation sessions
per evaluation item.}
\label{tab:bench_stats}
\begin{tabular}{lrrp{5.5cm}}
\toprule
Benchmark & Questions & Session depth & Query types \\
\midrule
LongMemEval~\cite{wu2024longmemeval}
  & 500  & Multi-session  & Single-session-preference, single-session-user, single-session-assistant, multi-session, temporal-reasoning, knowledge-update \\
LoCoMo~\cite{maharana2024locomo}
  & 1{,}986 & 600-turn peer conversations & Adversarial, multi-hop, temporal, preference; third-party named-entity queries \\
LongBench~\cite{bai2024longbench}
  & 1{,}750 & Single document & QA, summarization, multi-document reasoning (9 subtasks) \\
\midrule
\textbf{Total} & \textbf{4{,}236} & --- & --- \\
\bottomrule
\end{tabular}
\end{table}

\paragraph{LongMemEval.}
500 questions over long-term chat histories, testing six memory capabilities.
We use the oracle split (\texttt{longmemeval\_oracle.json}) which provides ground-truth
session annotations, allowing us to report oracle retrieval recall alongside end-to-end accuracy.

\paragraph{LoCoMo.}
1{,}986 questions over 600-turn conversations between two named speakers discussing
a shared social network.
Many questions ask about third-party people (e.g.\ ``What does Caroline do for work?'')
where the answer is embedded in narrative speech rather than direct statements.
This makes the benchmark particularly challenging for retrieval systems that rely on
surface-level keyword matching.

\paragraph{LongBench.}
1{,}750 questions across nine document-level subtasks from LongBench v1.
We include tasks spanning single-document QA (NarrativeQA, Qasper, MultiFieldQA-en),
multi-document QA (HotpotQA, 2WikiMultiHopQA, MuSiQue), and summarization
(GovReport, QMSum, MultiNews).
LongBench is qualitatively different from the other two benchmarks: it tests
document comprehension rather than episodic memory, and is included to evaluate
MemFlow's generalization beyond its primary session-based memory use case.

\subsection{GPT-4o-mini Judge Protocol}
\label{app:judge}

All systems are scored by a GPT-4o-mini judge~\cite{openai2024gpt4omini} that compares the predicted answer
against the gold reference and outputs a binary correct/incorrect judgment.

\paragraph{Judge prompts.}
The system and user prompts used for evaluation are:

\medskip
\noindent\textit{System:}
\begin{quote}\small
\textit{``You are an expert judge evaluating question-answering accuracy.
Be lenient on phrasing but strict on factual correctness.''}
\end{quote}

\noindent\textit{User template:}
\begin{verbatim}
Question: {question}
Gold Answer: {gold}
Predicted Answer: {predicted}

Is the predicted answer correct or semantically equivalent to the gold answer?
Accept partial matches if the key information is present.
Reply ONLY with JSON: {"correct": true} or {"correct": false}
\end{verbatim}

\paragraph{Why GPT-4o-mini judging over EM/F1.}
Exact match (EM) scores are near-zero across all systems (0.07\%--5.1\%),
while GPT-4o-mini accuracy ranges 9.6\%--74.3\%.
Token-level F1 (2.7\%--24.4\%) is also severely deflated.
This reflects the open-ended nature of memory QA: a correct answer such as
\textit{``GPS system malfunction on March 22, 2023''} has low lexical overlap with the
gold reference \textit{``GPS system not functioning correctly''} yet is clearly correct.
We report only GPT-4o-mini accuracy as the primary metric, consistent with
the evaluation protocol of LongMemEval~\cite{wu2024longmemeval}.

\paragraph{Choice of GPT-4o-mini as judge.}
We use GPT-4o-mini rather than GPT-4o to keep evaluation cost tractable
across 4{,}236 items $\times$ five systems.
On a held-out sample of 300 predictions, the two judges agree on 85.7\% of
items; in the 14.3\% of disagreements, GPT-4o-mini is the stricter judge
(GPT-4o accepts 35 answers that GPT-4o-mini rejects, vs.\ 8 in the reverse
direction).
Consequently, all reported accuracy figures are \emph{conservative} estimates
relative to the LongMemEval standard judge; replacing GPT-4o-mini with
GPT-4o would be expected to increase all system scores uniformly, leaving
relative rankings and \emph{pp}-gap claims unchanged.

\paragraph{Human audit.}
To further validate the automatic judge, we manually annotated 300
randomly sampled predictions stratified across benchmarks and systems.
GPT-4o-mini agreed with the human label on 89.6\% of items, with
disagreements concentrated in borderline semantic-equivalence cases. This
audit supports using GPT-4o-mini accuracy as a scalable semantic correctness
estimate rather than an exact lexical metric.

\paragraph{Fallback.}
In 0.9\% of cases where the OpenAI API is unavailable (network errors, rate limits),
we fall back to a token-overlap correctness check with a 50\% recall threshold.
These fallbacks are flagged in the output JSONL as \texttt{gpt4o\_fallback: true}.

\subsection{Baseline Implementation Details}
\label{app:baselines}

\paragraph{Direct QA (short, 2,333 tokens).}
The raw conversation history is truncated to 2{,}333 tokens and passed directly to
Qwen3-1.7B.
No retrieval or routing is performed.
The reported ``Answer Ctx'' for this baseline (avg.\ 2,458 tokens) exceeds 2,333 because it
includes the system prompt and question tokens appended to the truncated history; the
2,333-token figure refers strictly to the history truncation budget.
Token budget matches MemFlow's average packed context.

\paragraph{Direct QA (full, 9,800 tokens).}
Same as Direct QA (short) but truncated to 9{,}800 tokens.
Token budget matches MemFlow's full pipeline budget.

\paragraph{RAG.}
Hybrid BM25+dense retrieval (same retriever as MemFlow) with a fixed top-$k$ of 8.
No routing layer; all queries receive the same flat retrieval strategy.
Average packed context: 2{,}266 tokens.

\paragraph{ReAct~\cite{yao2023react}.}
Multi-step reasoning loop with access to the same hybrid BM25+dense retrieval
backend and deterministic tools used by MemFlow.
The SLM decides when to search, which tool call to issue, and when to answer.
Maximum 5 reasoning turns per query.
Average prompt tokens: 8{,}732. No validator or escalation.
This baseline tests whether an SLM can use the same retrieval and deterministic
tool interface through open-ended reasoning rather than MemFlow's bounded
route-then-compile execution policy.

\paragraph{MemGPT~\cite{packer2023memgpt}.}
Working memory + archival memory with model-driven page-in/page-out calls.
The SLM issues memory management commands (\texttt{recall\_memory},
\texttt{archival\_memory\_search}) as part of its response.
Average packed context: $\approx$1{,}260 tokens (estimated from retrieval metadata).

\paragraph{Memobase~\cite{chhikara2025mem0} (open-source Mem0 implementation).}
User-profile dual-store with BM25 retrieval.
Compresses all conversation history into a structured user profile;
retrieval operates over the profile rather than raw turns.
Average packed context: $\approx$260 tokens.
This extremely small context budget explains Memobase's low accuracy (9.6\%):
profile compression destroys episodic recall needed for most benchmark queries.

\paragraph{GPT-4o~\cite{openai2024gpt4o} (matched, 9,800 tokens).}
GPT-4o with the conversation history truncated to 9{,}800 tokens.
No retrieval or routing. Serves as an upper-bound reference at the same token budget
as MemFlow's full pipeline.

\paragraph{GPT-4o (ceiling, 120k tokens).}
GPT-4o with up to 120{,}000 tokens of conversation history.
Serves as the unconstrained frontier reference.

\section{Full Pipeline Algorithm}
\label{app:algorithm}

Algorithm~\ref{alg:memflow} provides pseudocode for the MemFlow read-path pipeline.

\begin{algorithm}[h]
\caption{MemFlow Read-Path Pipeline}
\label{alg:memflow}
\begin{algorithmic}[1]
\Require Query $q$, history $H$, SLM $\pi_\theta$
\Ensure Answer $a$
\State retries $\gets 0$

\If{token\_overlap$(q,\; \text{last-8 turns of } H) \geq 0.6$}
    \Return highest-overlap contiguous span from the last-8 turns \Comment{Stage 1}
\EndIf

\State $r \gets \textsc{RouterAgent}(q)$ \Comment{Stage 2: tag + routing flags}

\State $\text{evidence} \gets \begin{cases}
  \textsc{Tier1}(q) & \text{if } r.\text{tag} = \texttt{profile-injection} \\
  \textsc{Tier2}(q, H) & \text{if } r.\text{tag} = \texttt{targeted-extraction} \\
  \textsc{Tier3}(q, H, r) & \text{otherwise}
\end{cases}$ \Comment{Stage 3}

\State $C \gets \textsc{Pack}(\text{evidence},\; r.\text{tag})$ \Comment{Stage 4: ceiling 20{,}480 tok, avg 2{,}223}

\State $a \gets \pi_\theta\!\left(q,\, C,\, \textsc{TagPrompt}(r.\text{tag})\right)$ \Comment{Stage 5: answer agent}
\State tool\_rounds $\gets 0$
\While{tool call in $a$ \textbf{and} tool\_rounds $< 3$}
    \State $a \gets \pi_\theta(\textsc{ExecuteTool}(a))$ \Comment{at most 3 tool rounds}
\State tool\_rounds $\gets$ tool\_rounds $+ 1$
\EndWhile

\State $v \gets \textsc{Validate}(a, C, q)$ \Comment{Stage 6: grounding check}
\If{$\lnot v.\text{grounded}$ \textbf{and} retries $= 0$}
    \State retries $\gets 1$;\; $r \gets \textsc{Escalate}(r.\text{tag})$;\; \textbf{goto} Stage 3
\ElsIf{$\lnot v.\text{grounded}$}
    \Return \textit{``I could not find reliable information.''}
\EndIf
\State \Return $a$

\end{algorithmic}
\end{algorithm}

\section{Extended Experimental Results}
\label{app:extended}

\subsection{Router Accuracy: Per-Benchmark Breakdown}
\label{app:router_accuracy}

Table~\ref{tab:router_accuracy} reports triage routing accuracy across all three evaluation benchmarks.

\begin{table}[h]
\centering
\small
\caption{Router Agent \texttt{router\_correct} accuracy per benchmark.
87.7\% is from a dedicated router evaluation set (\texttt{scripts/eval\_router\_accuracy.py}).}
\label{tab:router_accuracy}
\begin{tabular}{lcc}
\toprule
Benchmark & $N$ & Routing accuracy \\
\midrule
Router eval set (dedicated) & ---    & \textbf{87.7\%} \\
LoCoMo (end-to-end)         & 1{,}986 & 88.6\% \\
LongBench (end-to-end)      & 1{,}750 & 90.9\% \\
LongMemEval (end-to-end)    & 500    & 72.4\% \\
\bottomrule
\end{tabular}
\end{table}

The three-tier cascade (rules $\to$ SLM $\to$ heuristic) ensures routing never fails outright.
LoCoMo (88.6\%) and LongBench (90.9\%) achieve higher end-to-end routing accuracy than the
dedicated router eval set (87.7\%), reflecting that their question structures map cleanly onto
MemFlow's intent-based action tags.
LongMemEval's lower accuracy (72.4\%) reflects the difficulty of aligning its question-type
ontology---which mixes topic and memory-type labels---with MemFlow's action tags; the
\texttt{router\_correct} labels for LongMemEval are derived from benchmark question-type
annotations rather than MemFlow-native ground truth.
The current router emits one mutually exclusive tag per query. This simplifies execution
and avoids open-ended tool plans, but it can under-serve compound questions that require
multiple simultaneous memory operations, such as temporal reasoning over facts that also
need conflict resolution.

\subsection{Per-Question-Type Results Across All Benchmarks}
\label{app:per_type}

\begin{figure}[h]
\centering
\includegraphics[width=\linewidth]{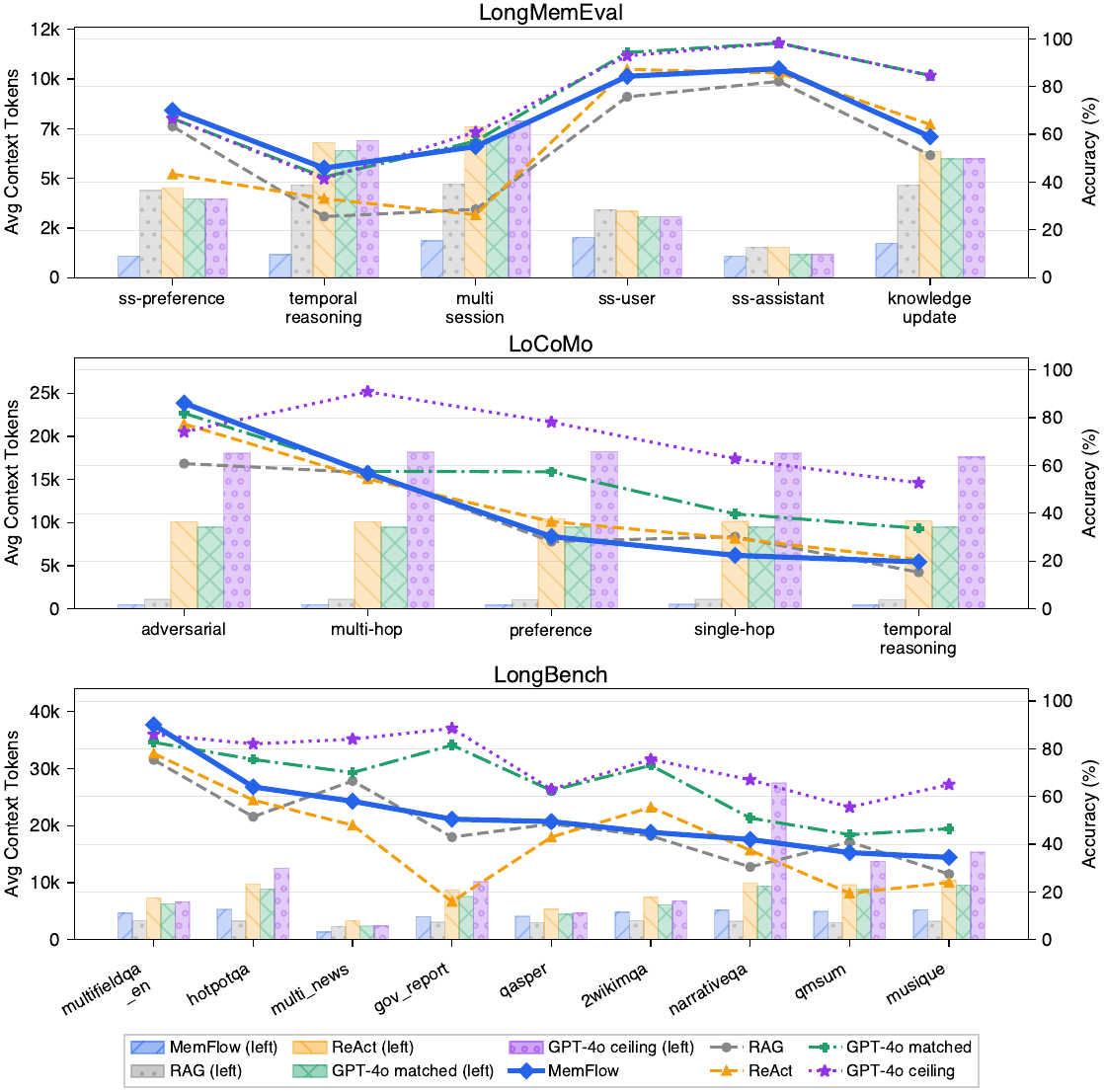}
\caption{Per-question-type accuracy (\%, lines, right axis) and Average context tokens (hatched bars, left
axis), grouped by benchmark. MemFlow (blue line), RAG (orange), ReAct
(green), and GPT-4o matched (red dashed) are shown for each question type.
Bold question-type labels indicate categories where MemFlow meets or
exceeds budget-matched GPT-4o. Packed context (hatched bars) reflects
the adaptive allocation per question type: LongBench tasks receive 4--5k
tokens on average due to the depth of evidence required; LoCoMo
adversarial queries require only $\sim$480 tokens. All accuracies judged
by GPT-4o-mini; $N{=}4{,}236$ across all three benchmarks.}
\label{fig:qtype_appendix}
\end{figure}

\begin{table}[h]
\centering
\small
\caption{Per-question-type accuracy (\%) and mean packed context tokens fed to the answer
agent across all three benchmarks.
Bold = MemFlow $\geq$ GPT-4o (matched). All accuracies judged by GPT-4o-mini.}
\label{tab:per_type_extended}
\begin{tabular}{lrrrrrrr}
\toprule
Question type & $n$ & MemFlow & RAG & ReAct & GPT-4o & GPT-4o & Avg pack \\
 & & & & & (matched) & (ceiling) & (tokens) \\
\midrule
\multicolumn{8}{l}{\textit{LongMemEval}} \\
\midrule
single-session-preference & 30  & \textbf{70.0} & 63.3 & 43.3 & 66.7 & 66.7 & 1{,}089 \\
temporal-reasoning        & 133 & \textbf{45.9} & 25.6 & 33.1 & 42.1 & 41.4 & 1{,}159 \\
multi-session             & 133 & 54.9 & 28.6 & 26.3 & 57.1 & 60.9 & 1{,}857 \\
single-session-user       & 70  & 84.3 & 75.7 & 87.1 & 94.3 & 92.9 & 2{,}031 \\
single-session-assistant  & 56  & 87.5 & 82.1 & 85.7 & 98.2 & 98.2 & 1{,}078 \\
knowledge-update          & 78  & 59.0 & 51.3 & 64.1 & 84.6 & 84.6 & 1{,}729 \\
\midrule
\textbf{Overall LongMemEval} & 500 & \textbf{61.8} & 46.0 & 50.2 & 67.8 & 68.4 & 1{,}543 \\
\midrule
\multicolumn{8}{l}{\textit{LoCoMo}} \\
\midrule
adversarial               & 446 & \textbf{86.1} & 60.8 & 77.4 & 81.8 & 74.0 & 482 \\
multi-hop                 & 841 & 56.7 & 57.0 & 54.3 & 57.6 & 90.8 & 473 \\
preference                & 96  & 30.2 & 28.1 & 36.5 & 57.3 & 78.1 & 415 \\
single-hop                & 282 & 22.3 & 30.1 & 29.4 & 39.7 & 62.8 & 517 \\
temporal-reasoning        & 321 & 19.6 & 15.3 & 20.6 & 33.6 & 52.6 & 412 \\
\midrule
\textbf{Overall LoCoMo}   & 1{,}986 & \textbf{51.2} & 45.9 & 49.6 & 56.6 & 76.3 & 469 \\
\midrule
\multicolumn{8}{l}{\textit{LongBench}} \\
\midrule
multifieldqa\_en          & 150 & \textbf{90.0} & 75.3 & 78.0 & 82.7 & 86.0 & 4{,}666 \\
hotpotqa                  & 200 & 64.0 & 51.5 & 58.5 & 75.5 & 82.0 & 5{,}321 \\
multi\_news               & 200 & 58.0 & 66.5 & 48.0 & 70.0 & 84.0 & 1{,}382 \\
gov\_report               & 200 & 50.5 & 43.0 & 16.0 & 81.5 & 88.5 & 4{,}036 \\
qasper                    & 200 & 49.5 & 48.5 & 43.0 & 62.5 & 63.0 & 4{,}120 \\
2wikimqa                  & 200 & 45.0 & 43.5 & 55.5 & 73.0 & 75.5 & 4{,}842 \\
narrativeqa               & 200 & 42.0 & 30.5 & 37.5 & 51.0 & 67.0 & 5{,}177 \\
qmsum                     & 200 & 36.5 & 41.0 & 19.5 & 44.0 & 55.5 & 4{,}962 \\
musique                   & 200 & 34.5 & 27.5 & 24.0 & 46.5 & 65.0 & 5{,}232 \\
\midrule
\textbf{Overall LongBench} & 1{,}750 & \textbf{51.1} & 46.7 & 41.2 & 64.7 & 73.7 & 4{,}408 \\
\bottomrule
\end{tabular}
\vspace{2pt}

{\footnotesize Bold = MemFlow $\geq$ GPT-4o (matched). See Section~\ref{app:per_type} for analysis.}
\end{table}

Figure~\ref{fig:qtype_appendix} and Table~\ref{tab:per_type_extended}
provide a fine-grained breakdown of MemFlow's performance across all
question types. Three patterns emerge.

\paragraph{Where MemFlow leads.}
On \textbf{LongMemEval}, MemFlow leads all SLM baselines on
\emph{single-session-preference} (70.0\%, $+$6.7\,pp over RAG) and
\emph{temporal-reasoning} (45.9\%, $+$20.3\,pp over RAG, $+$3.8\,pp over
GPT-4o matched) --- the two categories whose Tier-3 handlers
(chronological sort, date math, profile injection) are most directly
tailored to the query structure. On \textbf{LoCoMo}, MemFlow dominates
\emph{adversarial} queries (86.1\%, $+$8.7\,pp over ReAct and $+$4.3\,pp
over GPT-4o matched), where the Validator Agent's grounding check rejects
hallucinated answers before they are returned. On \textbf{LongBench},
MemFlow outperforms RAG on 7 of 9 task types, with the widest margins on
\emph{multifieldqa\_en} ($+$14.7\,pp) and \emph{gov\_report}
($+$7.5\,pp), where long-document comprehension benefits from structured
evidence extraction over raw truncation.

\paragraph{Remaining gaps.}
The largest gaps to GPT-4o matched are on \emph{knowledge-update}
($-$25.6\,pp on LongMemEval) and LoCoMo \emph{preference} ($-$27.1\,pp),
both of which require cross-session reconciliation of evolving facts that
the packer's compression budget cannot fully preserve. On
\emph{conflict-resolution}, router over-prediction (527 queries routed
vs.\ 78 ground-truth conflict queries) causes many non-conflict queries
to pass through a more expensive retrieval path, degrading accuracy and
representing the primary area for future routing improvement.

\paragraph{Context allocation reflects query complexity.}
The hatched bars in Figure~\ref{fig:qtype_appendix} show that MemFlow's
adaptive allocation distributes context budget in proportion to query
difficulty: LongBench tasks requiring multi-document evidence receive
4{,}000--5{,}300 tokens on average, while LoCoMo adversarial queries
require only $\sim$480 tokens. This confirms that the packer is not
simply filling a fixed budget but is responding to the structural
requirements of each query type.
\paragraph{Token budget vs.\ accuracy.}
The hatched bars in Figure~\ref{fig:qtype_appendix} should be read as evidence of
adaptive allocation, not as a monotonic predictor of accuracy.
On LongBench, all nine task types receive 1.3k--5.3k tokens---substantially more than
LoCoMo's 412--517 token budget---yet both achieve similar overall accuracy (LoCoMo 51.2\%,
LongBench 51.1\%), because LoCoMo's queries are typically answerable from a single
short span while LongBench requires aggregating evidence across long documents. Within a
benchmark, high context use can also mark intrinsically harder tasks such as multi-hop QA.

\subsection{Token Pipeline Cost Breakdown by Benchmark}
\label{app:token_breakdown_bench}

\begin{table}[h]
\centering
\small
\caption{Per-benchmark token statistics. ``Pack'' = tokens fed to the answer agent (packed context only).
``Pipeline'' = total tokens across all stages (router + answer + validator + escalation).}
\label{tab:bench_tokens}
\begin{tabular}{lrrrrr}
\toprule
Benchmark & $N$ & Pack mean & Pack median & Pipeline mean \\
\midrule
LongMemEval & 500     & 1{,}543 & 1{,}485 & 7{,}990  \\
LoCoMo      & 1{,}986 & 469     & 449     & 8{,}269  \\
LongBench   & 1{,}750 & 4{,}408 & 5{,}052 & 12{,}942 \\
\midrule
\textbf{Overall} & \textbf{4{,}236} & \textbf{2{,}223} & \textbf{874} & \textbf{10{,}167} \\
\bottomrule
\end{tabular}
\end{table}

The large variance across benchmarks reflects MemFlow's adaptive compression:
LoCoMo's peer-conversation queries tend to be answerable from a single short span
(median pack $<$500 tokens), while LongBench's document summarization tasks
require packing large portions of the document, approaching the 6{,}000-token Tier-2 ceiling.

\subsection{Escalation and Validator Analysis}
\label{app:escalation_analysis}

\paragraph{Escalation routing chains.}
When the Validator flags a response as ungrounded, the query is re-routed to a heavier
execution tier according to the per-tag escalation chains below.
Escalation is capped at one retry; if the escalated response also fails the validator,
the system returns a calibrated abstention.

\begin{table}[h]
\centering
\small
\caption{Escalation policy: one-retry tag-to-tag re-routing.}
\label{tab:escalation_policy}
\begin{tabular}{ll}
\toprule
Original tag & Retry tag \\
\midrule
\texttt{profile-injection}     & \texttt{targeted-extraction} \\
\texttt{targeted-extraction}   & \texttt{conflict-resolution} \\
\texttt{temporal-reasoning}    & \texttt{targeted-extraction} \\
\texttt{conflict-resolution}   & \texttt{targeted-extraction} \\
\texttt{broad-summarization}   & \texttt{targeted-extraction} \\
\texttt{constraint-validation} & \texttt{targeted-extraction} \\
\texttt{state-tracking}        & \texttt{conflict-resolution} \\
\bottomrule
\end{tabular}
\end{table}

\paragraph{Escalation statistics.}
Of the 4{,}236 evaluation queries:
\begin{itemize}[leftmargin=1.5em]
\item \textbf{98.0\%} of queries invoke the validator (4{,}150/4{,}236).
  The remaining 2.0\% are short-circuited by the Active Context Check before reaching
  the validator.
\item \textbf{14.9\%} of queries trigger escalation (validation failure + re-retrieval, 630/4{,}236).
  These represent cases where the initial answer agent response was flagged as ungrounded.
\item \textbf{3.4\%} of queries are truly escalated (146/4{,}236):
  the escalation re-retrieval produces a different response that is returned as the final answer.
  The remaining 11.5\% (484/4{,}236) are cases where escalation was triggered but the
  original response was ultimately retained or the re-retrieval failed to improve on it.
\end{itemize}
The 146 adopted escalations should not be read as the full contribution of the
``Escalation \& Validator'' ablation in Table~\ref{tab:ablation_bench}. That ablation removes
both the grounding gate and the retry path, so its effect includes prevented ungrounded
answers, abstention behavior, and retry opportunities, not only final-answer swaps.

\paragraph{Escalated query accuracy.}
Escalated queries (is\_escalated = True, $n=146$) achieve 10.3\% accuracy,
compared to 53.9\% for non-escalated queries.
This gap suggests that escalated queries are disproportionately drawn from
hard cases (knowledge-update, adversarial unanswerable) where the memory state
does not contain sufficient evidence regardless of retrieval strategy.
The escalation mechanism helps marginal cases through validation and retry, but cannot
recover fundamentally information-deficient queries.

\subsection{Ablation Study: Per-Benchmark Breakdown}
\label{app:ablation_bench}

\begin{table}[H]
\centering
\small
\caption{Per-benchmark ablation results. LME = LongMemEval, LCM = LoCoMo, LB = LongBench.
$\Delta$ = overall accuracy vs.\ full MemFlow (52.4\%).}
\label{tab:ablation_bench}
\begin{tabular}{lrrrrrr}
\toprule
Configuration & LME (\%) & LCM (\%) & LB (\%) & Overall (\%) & $\Delta$ \\
\midrule
MemFlow (full)         & 61.8 & 51.2 & 51.1 & 52.4 & --- \\
\midrule
-- Retrieval Strategy  & 45.6 & 33.9 & 30.1 & 33.7 & $-$18.7 \\
-- Uniform RAG         & 60.4 & 42.8 & 37.9 & 42.9 & $-$9.5  \\
-- Router              & 56.0 & 45.3 & 39.2 & 44.0 & $-$8.4  \\
-- Tools               & 60.8 & 45.1 & 37.9 & 44.0 & $-$8.4  \\
-- Escalation \& Validator & 58.2 & 44.3 & 41.2 & 44.7 & $-$7.7 \\
-- Packer              & 58.0 & 43.0 & 43.1 & 44.8 & $-$7.6  \\
\bottomrule
\end{tabular}
\end{table}

Table~\ref{tab:ablation_bench} extends the main-paper ablation (Table~3) with
per-benchmark accuracy for each configuration.
Removing the Retrieval Strategy causes the largest drop across all three benchmarks
($-$16.2 pp on LME, $-$17.3 pp on LCM, $-$21.0 pp on LB), confirming that adaptive
intent-aware retrieval is the primary driver of MemFlow's advantage regardless of benchmark type.
Removing Uniform RAG and the Router produce comparable overall drops ($-$9.5 and $-$8.4 pp),
but LongBench is disproportionately affected by both Retrieval Strategy and Tools,
consistent with its document-level tasks requiring broader chunk coverage and arithmetic reasoning.
The Packer and Escalation \& Validator components show the smallest individual drops
($-$7.6 and $-$7.7 pp overall), yet both contribute consistently across all three benchmarks.

\section{Broader Impact}
\label{app:broader_impact}

MemFlow is a training-free memory scaffolding system designed to enable
sub-3B-parameter SLMs to operate as capable long-horizon agents.
MemFlow's deterministic, modular design lowers the barrier to deploying capable AI agents
in resource-constrained environments: on-device mobile assistants, edge-computing systems,
and institutions with limited GPU infrastructure.
By preserving a bounded context budget,
MemFlow makes it practical to run persistent personal assistants locally
rather than relying on cloud-hosted frontier models, with potential privacy benefits
for users who prefer to keep their conversation history on-device.
The architecture is model-agnostic: it improves all five tested SLMs,
suggesting broad applicability across the open-source model ecosystem.

\paragraph{Risks and limitations.}
Persistent memory systems maintain long-term user profiles derived from conversation history.
If deployed without explicit user consent or adequate data governance,
such systems could enable unauthorized behavioral profiling, preference tracking,
or manipulation.
Users should be clearly informed of what is stored, how it is used, and how to delete it. As with all current SLM-based memory systems, accuracy remains below the threshold
required for high-stakes decisions (medical, legal, financial) without human oversight.
The reliance on a GPT-4o-mini judge for evaluation introduces a potential evaluation bias:
judge scores may not perfectly align with human judgments, particularly for
ambiguous or culturally specific answers.
All experiments are conducted on English-language datasets;
generalization to other languages has not been tested.

\end{document}